\documentclass[12pt]{article} 

\usepackage{float}
 \usepackage{amsmath} 
  \usepackage{amssymb}   
\usepackage{amsthm} 
\usepackage{geometry}
\usepackage{graphicx} 
\usepackage{multirow} 
\usepackage{cite}
\usepackage{color,xspace}
\usepackage{extarrows}

\numberwithin{equation}{section}
\DeclareSymbolFont{extraup}{U}{zavm}{m}{n}
\DeclareMathSymbol{\vardiamond}{\mathalpha}{extraup}{87}
\DeclareMathSymbol{\varheartsuit}{\mathalpha}{extraup}{86}
\usepackage{float}
\restylefloat{table}
\allowdisplaybreaks

\newcommand{\beq}{\begin{equation}}
\newcommand{\eeq}{\end{equation}}
\newcommand{\bea}{\begin{eqnarray}}
\newcommand{\eea}{\end{eqnarray}}

\newcommand{\gsim}{\lower.7ex\hbox{$\;\stackrel{\textstyle>}{\sim}\;$}}
\newcommand{\lsim}{\lower.7ex\hbox{$\;\stackrel{\textstyle<}{\sim}\;$}}

\newcommand{\be}{\begin{equation}}
\newcommand{\ee}{\end{equation}}
\newcommand{\ba}{\begin{eqnarray}}
\newcommand{\ea}{\end{eqnarray}}

\newcommand{\nn}{\nonumber}

\title{Stochastic gravitational waves from spin-3/2 fields -- Hunting SUSY in the sky}

\date{}

\begin{document}
\begin{flushright}
\end{flushright}
\begin{center}

\vspace{1cm}
{\Large{\bf Stochastic gravitational waves from spin-3/2 fields \\
Hunting SUSY in the sky}}

\vspace{1cm}

\textbf{Karim Benakli$^\varheartsuit$ \let\thefootnote\relax\footnote{$^\varheartsuit$kbenakli@lpthe.jussieu.fr}
Yifan Chen$^{\spadesuit}$ \let\thefootnote\relax\footnote{$^\spadesuit$yifan.chen@lpthe.jussieu.fr}
Peng Cheng$^{\vardiamond}$ \let\thefootnote\relax\footnote{$^\vardiamond$peng.cheng@polytechnique.edu}
 Ga\"etan~Lafforgue-Marmet$^\clubsuit$ \footnote{$^\clubsuit$glm@lpthe.jussieu.fr}
\\[5mm]}
 
 
{ {}$^{\varheartsuit \spadesuit \clubsuit}$ \sl Laboratoire de Physique Th\'eorique et Hautes Energies (LPTHE),\\ UMR 7589,
Sorbonne Universit\'e et CNRS, 4 place Jussieu, 75252 Paris Cedex 05, France.}

{ {}$^{\vardiamond}$ \sl Centre de Physique Th\'eorique, \'Ecole Polytechnique, CNRS,\\
Universit\'e Paris-Saclay, 91128 Palaiseau Cedex, France}

\end{center}
\vspace{0.7cm}

\abstract{Stochastic gravitational waves can be produced during the preheating when out-of-equilibrium particles are produced with an anisotropic stress-tensor. We discuss the case where these particles carry spin 3/2. We compute the spectrum of the gravitational waves generated by the transverse and longitudinal components. We find a different scaling of the spectrum near the peak and the longitudinal components lead to an enhancement when compared to spin-1/2 fermions with Yukawa couplings. We note, as expected, that the corresponding typical frequency is too high for the current observation and calls for ultra-high frequency gravitational wave detectors in the future.}

\newpage
\setcounter{footnote}{0}

\section{Introduction}

The detection of gravitational waves from astrophysical sources \cite{Abbott:2016blz} is another successful test of general relativity \cite{Weinberg:1972kfs}. It gives rise to a growing interest in new possible  gravitational wave detectors with also the hope of discovery of new sources that may be of cosmological nature, for example \cite{Starobinsky:1979ty}. While an apparatus like  \cite{Audley:2017drz} can detect such corresponding  gravitational waves at low frequency,  there are cosmological sources that lead to signals at higher frequencies. Examples are gravitational waves produced during the preheating era \cite{Traschen:1990sw}. New experiments need to be designed to detect them (e.g., \cite{Sabin:2014bua}). 

Here, we are interested in systems of quantum gases with large anisotropic stress tensor that produce stochastic gravitational waves \cite{Ford:1982wu}. These  are expected to be present during preheating when particles are violently produced far from thermal equilibrium  \cite{Khlebnikov:1997di}. Previous studies have focused on bosonic sources including both scalars and gauge bosons \cite{Easther:2006gt, Dufaux:2007pt} as well as Dirac spin-1/2 fermions \cite{Enqvist:2012im}. In this work, we study the  gravitational wave signals coming from nonadiabatic spin-3/2 gases.

No fundamental spin-3/2 particles are known in nature, but composite states are produced as hadrons. It is not easy to write a consistent Lagrangian for fundamental spin 3/2 with minimal coupling to a gauge boson that would would make them strongly interacting with known particles and allow them to be detected easily. A major difficulty is the Velo-Zwangizer problem: when minimally coupled to electromagnetism, the longitudinal mode, i.e., helicity $\pm \frac{1}{2}$-states, have nonhyperbolic equations of motion that lead to noncausal propagation \cite{Velo:1969bt}. One is then led to either consider nonminimal couplings or spin-3/2 states with gravitational interactions. In the latter case, it is well known that the longitudinal mode can be understood as due to a super-Higgs mechanism. In a nonunitary gauge and in the supersymmetric phase, the longitudinal mode is a well-behaved fermion that has a causal behavior and therefore the Velo-Zwanziger problem is not expected to show up in supergravity and string theory, for example \cite{Porrati:2010hm, Rahman:2011ik}.

As our main candidate for a fundamental spin-3/2 particle is the gravitino that has only gravitational interactions, it is important to search for experiments and observations that are sensitive to such interactions. Here, we investigate the features: shape, amplitude, and peak frequency of gravitational waves that are produced during preheating. Our aim is to compare them to other signals due to particles with different spins in order to investigate if the signals can, in principle, be distinguished if experiments become sensitive to the expected frequencies and amplitudes in the future. Looking for signatures of spin-3/2 fundamental particles in the early Universe is another way to search for supersymmetry signatures. The raised question is also the following: if supersymmetry does not show up at collider experiments, could it still be an important ingredient of early Universe cosmology? Here we study one way it could affect the production of gravitational waves.

Section 2 reviews the basics facts about Rarita-Schwinger description of spin-3/2 fields. Section 3 presents the computation of the spectrum of energy density of gravitational waves per frequency interval. It contains the main results of this work: the master formula for the estimate of produced  gravitational waves from both the transverse and longitudinal modes. We show an enhancement of the latter compared to expected signal from spin-1/2 fermions. A quantitative evaluation requires explicit examples where wave functions of the produced spin-3/2  states can be computed. We illustrate our results in a simple model in section 5. Our results are briefly summarized in the conclusions.

\newpage

\section{The Rarita-Schwinger fields}

In order to describe a spin-3/2 field, one starts with a $\psi_\mu$ in the spinor-vector representation of the Lorentz group that obeys a Dirac equation. Using representations of $SU(2)_L \times SU(2)_R$ for Weyl spinors, this is obtained from the tensor product  of spin representations as
\begin{align}
(\frac{1}{2},\frac{1}{2}) \otimes (\frac{1}{2},0) =  \frac{1}{2} \oplus (1 \otimes \frac{1}{2}) = \frac{1}{2} \oplus \frac{1}{2} \oplus \frac{3}{2} \ ,
\label{decompose}
\end{align}
and two constraints have to be imposed in order to project out the two additional spin-1/2 representations. In flat space time, the spin-3/2 field $\psi_\mu$ obeys then the equations
\ba
(i\partial \!\!\!/ - m_{3/2}) \psi_\mu &=& 0,  \label{eom}\\
 \gamma^\mu \psi_\mu &=& 0,\label{constraint1}\\
\partial^\mu \psi_\mu &=& 0. \label{constraint2}
\label{equations of motion3}\ea
These can be obtained from the Rarita-Schwinger Lagrangian
\be
 \mathcal{L} = -\frac{1}{2}\epsilon^{\mu\nu\rho\sigma}\bar{\psi}_\mu \gamma_5\gamma_\nu\partial_\rho\psi_\sigma - \frac{1}{4} m_{3/2} \bar{\psi}_\mu [\gamma^\mu, \gamma^\nu] \psi_\nu \ .
 \label{LRS}
 \ee
In the momentum space, the solution $\tilde{\psi}^\mu$ to the above equations of motion reads:
\be 
\tilde{\psi}^\mu_{\textbf{p}, \lambda} = \sum_{s = \pm1, l = \pm1, 0} \langle 1, \frac{1}{2}, l, \frac{s}{2} | \frac{3}{2}, \lambda \rangle u_{\textbf{p}, \frac{s}{2}}\epsilon^\mu_{\textbf{p}, l},
\label{solRS}
\ee
where $  \langle 1, \frac{1}{2}, l, \frac{s}{2} | \frac{3}{2}, \lambda \rangle$ are the Clebsch-Gordan coefficients in the decomposition (\ref{decompose}) in the standard notation. The $\epsilon^\mu_{\textbf{p}, l}$ and $u_{\textbf{p}, \frac{s}{2}}$ are normalized solutions of massive spin-1 and spin-1/2 fields equations.
Explicit expressions of the decomposition can be  found in \cite{Auvil:1966eao,Moroi:1995fs,Benakli:2014bpa}.
Using the identity
\be 
\epsilon^{\mu\nu\rho\sigma} \gamma_5 \gamma_\sigma = -i\gamma^{[\mu\nu\rho]}, 
\ee
the Lagrangian (\ref{LRS}) can be written as
\ba \mathcal{L} &=& \frac{1}{2} \bar{\psi}_\mu (i\partial \!\!\!/ - m_{3/2}) \psi^\mu -\frac{i}{2} \bar{\psi}_\mu (\gamma^\mu \partial^\nu + \gamma^\nu\partial^\mu) \psi_\nu\nn\\
 &+& \frac{i}{2} \bar{\psi}_\mu\gamma^\mu\partial \!\!\!/\gamma^\nu\psi_\nu + \frac{1}{2}m_{3/2}\bar{\psi}_\mu \gamma^\mu\gamma^\nu\psi_\nu. \ea
The symmetric stress tensor can be derived as
\ba 
T_{\alpha\beta} &=& \frac{e_{c\alpha}}{2e} \frac{\delta (e \mathcal{L})}{\delta e^\beta_c} + (\alpha \leftrightarrow \beta)\nn\\
&=& \frac{i}{4} \bar{\psi}_\mu \gamma_{(\alpha}\partial_{\beta)}\psi^\mu - \frac{i}{4}\bar{\psi}_\mu \gamma_{(\alpha} \partial^\mu \psi_{\beta)} + h.c.,\label{stresstensor3/2}
\ea
where in the last step we used the equations of motion and constraints (\ref{eom}) - (\ref{constraint2}) to eliminate irrelevant terms. 
The Majorana spin-3/2 fields are quantized,
\be 
\psi^\mu(\textbf{x}, t) = \sum_{\lambda =\pm \frac{3}{2}, \pm \frac{1}{2}} \int \frac{d\textbf{p}}{(2\pi)^3} e^{-i\textbf{p}\cdot\textbf{x}} \{ \hat{a}_{\textbf{p}, \lambda} \tilde{\psi}^\mu_{\textbf{p}, \lambda}( t) + \hat{a}_{-\textbf{p}, \lambda}^\dagger \tilde{\psi}^{\mu C}_{\textbf{p}, \lambda}( t) \},
\label{spin-3/2 statequantisation}
\ee
where the annihilation and creation operators are time independent and satisfy
\be 
\{\hat{a}_{\textbf{p}, \lambda}, \hat{a}_{\textbf{p}', \lambda'}^\dagger\} = (2\pi)^3 \delta_{\lambda, \lambda'} \delta^{(3)} (\textbf{p} - \textbf{p}'). 
\ee

\section{Gravitational wave production}

Here, we compute the spectrum of energy density of gravitational waves produced by a gas of spin-$3/2$ states.  We consider wavelengths in the subhorizon limit, the effects of curvature and torsion can be neglected. 

The gravitational waves can be described as linear tensor perturbations, here in the transverse-traceless (TT) gauge, of the Friedman-Robertson-Walker (FRW) metric,
\be 
ds^2 = a^2(\tau) [-d\tau^2 + (\delta_{ij} + h_{ij}) dx^i dx^j],
\label{FRW}
\ee
 where $\tau$ is the conformal time. The linear perturbation part of Einstein equations leads to the  gravitational wave equations of motion,
 \be 
 \ddot{h}_{ij} + 2\mathcal{H} \dot{h}_{ij} - \nabla h_{ij} = 16\pi G\Pi_{ij}^{TT},
 \ee 
 where the dot (.) stands for the derivative with respect to the conformal time $\tau$, $\mathcal{H} = \frac{\dot{a}}{a}$ is then the comoving Hubble rate, and $\Pi_{ij}^{TT}$ is the TT part of the anisotropic stress tensor. In order to avoid manipulating the nonlocal projection operator in configuration space, we perform a Fourier transform of the stress tensor $T_{\mu \nu}$ in terms of comoving wave number $\textbf{k}$. Then $-\nabla$ gives $k^2 = |\textbf{k}|^2$ and we can write
 \be
  \Pi_{ij}^{TT} (\textbf{k}, t) = \Lambda_{ij, lm} (\hat{\textbf{k}}) (T^{lm}(\textbf{k}, t) - \mathcal{P} g^{lm} ),
  \label{TTstress}
  \ee
 where $\mathcal{P}$ is the background pressure and $\Lambda_{ij, lm} $ is the TT projection tensor:
 \be  
 \Lambda_{ij, lm} (\hat{\textbf{k}}) \equiv P_{il} (\hat{\textbf{k}}) P_{jm} (\hat{\textbf{k}}) - \frac{1}{2} P_{ij} (\hat{\textbf{k}}) P_{lm} (\hat{\textbf{k}}), \qquad P_{ij} (\hat{\textbf{k}}) = \delta_{ij} - \hat{\textbf{k}}_i \hat{\textbf{k}}_j. 
 \label{defkproj}
 \ee

We assume the stochastic gravitational background to be isotropic, stationary, and Gaussian, therefore completely specified by its power spectrum.
For the subhorizon modes $k \gg \mathcal{H}$, the spectrum of energy density per logarithmic frequency interval can be written as \cite{Dufaux:2007pt}
\be 
\frac{d\rho_{ GW}}{d \textrm{log} k} (k, t) = \frac{2 G k^3}{\pi a^4 (t)}\int_{t_I}^t dt' \int_{t_I}^t dt'' a(t') a(t'') {\cos}[k (t' - t'')] \Pi^2 (k, t', t''),
\label{ gravitational wavespectrum}
\ee
where $\Pi^2 (k, t', t'')$ is the unequal-time correlator of $\Pi_{ij}^{TT}$ defined as
\be 
\langle \Pi_{ij}^{TT} (\textbf{k}, t) \Pi^{TT ij} (\textbf{k}', t') \rangle \equiv (2\pi)^3 \Pi^2 (k, t, t') \delta^{(3)} (\textbf{k} - \textbf{k}'),
\label{UTC}
\ee
and $\langle ... \rangle$ denotes ensemble average. 

To make Eq. (\ref{ gravitational wavespectrum}) from massive particles nonzero, we expect the time dependence of the wave function to vary nonadiabatically with frequencies which we discuss in the next section. We restrict to a situation where $m_{3/2} \gg \mathcal{H}$ in which case we can use flat limit quantization. We choose to parametrize the time dependence by writing the spinor wave functions  as functions of time  while keeping the vector polarizations $\epsilon^\mu_{\textbf{p}, l}$ constant,
\be 
\tilde{\psi}^{\mu}_{\textbf{p}, \lambda}  (t) =  \sum_{s = \pm 1, l = \pm 1, 0} \langle 1, \frac{1}{2}, l, \frac{s}{2} | \frac{3}{2}, \lambda \rangle \epsilon^{\mu}_{\textbf{p}, l} \mathbf{u}^{(|\lambda|)}_{\textbf{p}, \frac{s}{2}} (t),
\label{solRS2}
\ee
where we defined
\be 
\mathbf{u}^{(|\lambda|) T}_{\textbf{p}, \frac{s}{2}} (t) = (u_{\textbf{p}, +}^{(|\lambda|)}(t) \chi_s^T(\textbf{p}) ,  \, \,  s \,  u_{\textbf{p}, -}^{(|\lambda|)} (t)  \chi_s^T (\textbf{p})),
 \ee
expressed in terms of the (scalar) wave function $u_{\textbf{p}, \pm}^{(|\lambda|)}(t)$ and the two-component normalized eigenvectors $\chi_s (\textbf{p})$ of the helicity operator. 
 
We first consider the Hamiltonian of the fields, which is the space integral of the $T^{00}$ component of the stress tensor (\ref{stresstensor3/2}),
\ba H(t) &=& \int d\textbf{x}\, T^{00} (\textbf{x}, t)\nn\\
&=& \int d\textbf{x}\, \frac{i}{4} \bar{\psi}_\mu (\textbf{x}, t) \gamma^0 \partial_t \psi^\mu (\textbf{x}, t)\label{H2} + h.c. \nn\\
&=&  \int d\textbf{x}\,  \frac{i}{4} \bar{\psi}^{(\frac{1}{2})} (\textbf{x}, t) \gamma^0 \partial_t \psi^{(\frac{1}{2})}  (\textbf{x}, t) + \frac{i}{4}\bar{\psi}^{(\frac{3}{2})} (\textbf{x}, t) \gamma^0 \partial_t \psi^{(\frac{3}{2})}  (\textbf{x}, t) + h.c., \nn\\
\label{H3}\ea
where in the second line the second term of Eq. (\ref{stresstensor3/2}) vanishes since we can do the integral by part  leading to the constraint Eq. (\ref{constraint2}). In the last line, we used the property $\epsilon^{\mu}_{\textbf{p}, l} \epsilon^*_{\mu \textbf{p}, l'} = \delta_{l, l'}$ and $\chi_s^\dagger (\textbf{p}) \chi_{s'} (\textbf{p}) = \delta_{s, s'}$. The two spinors are defined as
\be 
\psi^{(|\lambda|) } (\textbf{x}, t) = \sum_{s =\pm 1} \int \frac{d\textbf{p}}{(2\pi)^3} e^{-i\textbf{p}\cdot\textbf{x}} \{\hat{a}_{\textbf{p}, \lambda} \mathbf{u}^{(|\lambda|)}_{\textbf{p}, \frac{s}{2}} (t) + \hat{a}_{-\textbf{p}, \lambda}^\dagger {\mathbf{v}}^{(|\lambda|)}_{\textbf{p}, \frac{s}{2}} (t) \}.
\label{spinor quantization}
\ee
Substituting Eq. (\ref{spinor quantization}) into Eq. (\ref{H3}) does not give a diagonal form in terms of annihilation and creation operators. Thus we need to do the Bogoliubov transformation
\ba \hat{\tilde{a}}_{\textbf{p}, \lambda} (t) &=& \alpha_{\textbf{p}}^{(|\lambda|) } (t)  \, \hat{a}_{\textbf{p}, \lambda} + \beta_{\textbf{p}}^{(|\lambda|) } (t) \, \hat{a}^\dagger_{-\textbf{p}, \lambda},\nn
\ea
to make the Hamiltonian  (\ref{H3}) diagonal,
\be H(t) = \int \frac{d\textbf{p}}{(2\pi)^3}  \sqrt{m_{3/2}^2 + p^2} \sum_{\lambda = \pm \frac{1}{2}, \pm \frac{3}{2}}\hat{\tilde{a}}_{\textbf{p}, \lambda}^\dagger (t) \, \hat{\tilde{a}}_{\textbf{p}, \lambda} (t),\label{Hnewbasis}\ee
where $p = |\textbf{p}|$ and $\alpha_{\textbf{p}}^{(|\lambda|) } (t)$, $\beta_{\textbf{p}}^{(|\lambda|) } (t)$ are complex numbers satisfying $| \alpha_{\textbf{p}}^{(|\lambda|) } (t) |^2 + |\beta_{\textbf{p}}^{(|\lambda|) } (t) |^2 = 1$. In the Heisenberg picture, the expectation value is defined by projecting the time-dependent operator on the initial vacuum $|0\rangle$ that corresponds to vanishing number density. Using $n^{(\lambda)}_\textbf{p}(t) = \hat{\tilde{a}}_{\textbf{p}, \lambda}^\dagger (t) \, \hat{\tilde{a}}_{\textbf{p}, \lambda} (t)$  leads to the occupation number
\ba \langle 0 | n^{(\lambda)}_\textbf{p}(t) | 0 \rangle &=& |\beta_{\textbf{p}}^{(|\lambda|) } (t) |^2\nn\\
  &=& \frac{ \sqrt{m_{3/2}^2 + p^2} - p \, \textrm{Re} (u_{\textbf{p}, +}^{(|\lambda|) *}(t) \, u_{\textbf{p}, -}^{(|\lambda|)}(t)) - m_{3/2}\, (1 - |u_{\textbf{p}, +}^{(|\lambda|)}(t)|^2)}{2 \sqrt{m_{3/2}^2 + p^2}}.\nn\\
  \label{occupationnumber}\ea
  
We also get a time-dependent physical vacuum satisfying
\be \hat{\tilde{a}}_{\textbf{p}, \lambda} (t) |0_t\rangle = 0. \label{timedependentvacuum}\ee

We next consider the sources of the gravitational waves. Plugging the mode decomposition (\ref{spin-3/2 statequantisation}) into Eq. (\ref{TTstress}) leads to
\ba 
\Pi_{ij}^{TT} (\textbf{k}, t) &=& \frac{1}{4}\Lambda_{ij, lm} \int \frac{d\textbf{p}}{(2\pi)^3} \{ \hat \Pi^{lm} (\textbf{p}, t) + h.c. \},
\label{PiijTT2}
\ea
where $\textbf{k}$ is the momentum mode of the gravitational wave and
\ba 
{\hat \Pi^{lm}}(\textbf{p}, t)&=& \left[ \hat{a}_{-\textbf{p}, \lambda} \bar{\tilde{\psi}}^{\mu C}_{\textbf{p}, \lambda} + \hat{a}^\dagger_{\textbf{p}, \lambda} \bar{\tilde{\psi}}^\mu_{\textbf{p}, \lambda}\right] \gamma^{(l}\partial^{m)}
 \left[ \hat{a}_{\textbf{p} + \textbf{k}, \lambda'} \tilde{\psi}_{\mu\textbf{p}+\textbf{k}, \lambda'} + \hat{a}^\dagger_{-\textbf{p}-\textbf{k}, \lambda'} \tilde{\psi}^{C}_{\mu \textbf{p}+\textbf{k}, \lambda'}\right] \nn \\
&& \! \!  \! \!  \! \! -  \left[ \hat{a}_{-\textbf{p}, \lambda} \bar{\tilde{\psi}}^{\mu C}_{\textbf{p}, \lambda} + \hat{a}^\dagger_{\textbf{p}, \lambda} \bar{\tilde{\psi}}^\mu_{\textbf{p}, \lambda}\right]  \gamma^{((l}\partial_\mu  \left[\hat{a}_{\textbf{p} + \textbf{k}, \lambda'} \tilde{\psi}^{m)}_{\textbf{p}+\textbf{k}, \lambda'} + \hat{a}^\dagger_{-\textbf{p}-\textbf{k}, \lambda'} \tilde{\psi}^{m) C}_{\textbf{p}+\textbf{k}, \lambda'}\right]. \nn \\
\label{PiijTT22}
\ea

Notice that (\ref{defkproj}) implies that $\Lambda_{ij, lm} k_l = \Lambda_{ij, lm} k_m = 0$, which removes the linear $k$ dependence from $\partial_m$ in the first line of Eq. (\ref{PiijTT22}), similar to the case of scalars or spin-1/2 fermions \cite{Enqvist:2012im}. However, in the second line of Eq. (\ref{PiijTT22}), $\partial_\mu$ leads to nonvanishing $k_\mu$ contracting with $\epsilon^\mu_{\textbf{p}, m}$, which is an important property of spin-3/2 gases.

The annihilation and creation operators lead to $2^4 = 16$ combinations among which only one contributes to nontrivial results:
\ba
&&  \! \!  \! \!  \! \! \! \!  \! \!  \! \! \! \!  \! \!  \! \! \! \!  \! \!  \! \! \! \!  \! \!  \!    \langle 0| \hat{a}_{-\textbf{p}, \lambda} \hat{a}_{\textbf{k} + \textbf{p}, \kappa} \hat{a}^\dagger_{\textbf{q}, \lambda'} \hat{a}^\dagger_{\textbf{k}'-\textbf{q}, \kappa'} |0 \rangle =  \nn\\
&& (2\pi)^6 \delta^{(3)} (\textbf{k} - \textbf{k}') \{ \delta^{(3)} (\textbf{k} + \textbf{p} -\textbf{q}) \delta_{\lambda, \kappa'} \delta_{\kappa, \lambda'}
-  \delta^{(3)} (\textbf{p} + \textbf{q}) \delta_{\lambda, \lambda'} \delta_{\kappa, \kappa'}\}.
\label{PiPivev} 
\ea
The two terms inside the brackets in (\ref{PiPivev}) come from the Majorana nature, assumed for the spin-3/2 fields, and lead to the same results. 

It is convenient to define
\be 
\textbf{p}'= \textbf{p} + \textbf{k}.
\ee

We now turn to the unequal-time correlator and write it in terms of 4-spinors,
\be 
\Pi^2 (k, t, t') = 2\int \frac{d\textbf{p}}{(2\pi)^3}  \left[ \bar{\textbf{v}}^{(|\lambda|)}_{\textbf{p}, \frac{s}{2}}(t) \Delta_{ij}^{\lambda s, \lambda's'} (t) \textbf{u}^{(|\lambda'|)}_{\textbf{p}', \frac{s'}{2}} (t)\right]   \left[ \bar{\textbf{u}}^{(|\lambda'|)}_{\textbf{p}', \frac{r'}{2}} (t') \Delta_{ij}^{\lambda r, \lambda'r'} (t')^* \textbf{v}^{(|\lambda|)}_{\textbf{p}, \frac{r}{2}} (t')\right],
\label{UTC4spinors}
\ee
where $ \textbf{v}^{(|\lambda|)}_{\textbf{p}, \frac{r}{2}} =  i\gamma^0\gamma^2 \bar{\textbf{u}}^{|\lambda| T}_{\textbf{p}, \frac{r}{2}}$ and
\ba 
\Delta_{ij}^{\lambda s, \lambda's'} (t) &=&  \frac{1}{4}\Lambda_{ij, lm}\langle 1, \frac{1}{2}, r, \frac{s}{2} | \frac{3}{2}, \lambda \rangle \langle 1, \frac{1}{2}, r', \frac{s'}{2} | \frac{3}{2}, \lambda' \rangle \times \nonumber\\
&&\{ 2\epsilon_{\mu\textbf{p}, r} \epsilon^{\mu}_{\textbf{p}', r'}  \, \, p^{(l} \gamma^{m)} - \epsilon_{\mu\textbf{p}, r}p'^\mu \epsilon^{(l}_{\textbf{p}', r'}\gamma^{m)} - \epsilon_{\mu\textbf{p}', r'} p^\mu \epsilon^{(l}_{\textbf{p}, r}\gamma^{m)} \}.\nn\\
\label{Deltaij}
\ea

We separate the calculation into two parts, $\lambda, \lambda' = \pm \frac{3}{2}$ and $\lambda, \lambda' = \pm \frac{1}{2}$, since the gravitational waves considered are produced mainly by relativistic states and the different helicity states in general are produced differently (see e.g., for gravitinos \cite{Kallosh:1999jj, Giudice:1999yt}).

We first consider the case of $\lambda, \lambda' = \pm\frac{3}{2}$. Such a restriction can be thought as working in the massless limit for the spin-3/2  state. For a gravitino, this is the high energy limit before the spontaneous breaking of supersymmetry. The mode decomposition (\ref{solRS2}) reads
\be 
\tilde{\psi}_{\textbf{p}, \pm \frac{3}{2}}^{\mu } (t) = \epsilon^{\mu }_{\textbf{p}, \pm1}\, \mathbf{u}^{({3}/{2})}_{\textbf{p},\pm \frac{1}{2}} (t).
\ee
We calculate the corresponding unequal-time correlator (\ref{UTC4spinors}),
\ba 
\Pi^2_{\frac{3}{2}} (k, t, t') &=&\frac{1}{32\pi^2} \int dp\, d\theta \, \, K^{(\frac{3}{2})} (p, k, \theta, m_{3/2}) \, \, \, \,  W^{(\frac{3}{2})}_{\textbf{k}, \textbf{p}} (t) W^{(\frac{3}{2})*}_{\textbf{k}, \textbf{p}} (t'),
\label{UTC3/2}
\ea
where $\theta (\theta')$ is the angle between $\textbf{k}$ and $\textbf{p} (\textbf{p}')$ and 
\ba 
K^{(\frac{3}{2})} (p, k, \theta, m_{3/2}) &= & p^2 k^2 \{5 \sin^3\theta  \sin^2\theta' +  \sin^2(\theta - \theta')\sin\theta\}
+ 4p^4 \sin^4\theta \sin\theta'.\nn\\
 \label{defK32}
 \ea
There is no final dependence on $p'$ and $\theta'$ as these are expressed  before integration as functions of $p$, $k$, and $\theta$,
\be 
p' = \sqrt{p^2 + k^2 +2 k p \ {\cos} \theta}, \qquad \theta' = \textrm{arccos} (\frac{p\ {\cos}\theta + k}{ \sqrt{p^2 + k^2 +2 k p\ {\cos} \theta}}).
\ee
We also defined
\be
 W^{(|\lambda|)}_{\textbf{k}, \textbf{p}} (t) = u_{\textbf{p}, +}^{(|\lambda|)}(t) u_{\textbf{p}', +}^{(|\lambda|)}(t) - u_{\textbf{p}, -}^{(|\lambda|)}(t) u_{\textbf{p}', -}^{(|\lambda|)}(t)
 \label{Wdef}
 \ee
to isolate kinematical factors from parts containing the wave functions. 

An important step before extracting a quantitative result is to remove the ultraviolet divergence in the momentum integral of the unequal-time correlator  (\ref{UTC3/2}). The regularized operator is built from the nonregularized one by subtracting the zero point fluctuations. Since at each time $t$ the physical vacuum is different, we should use the time-dependent vacuum defined in Eq. (\ref{timedependentvacuum}),
\ba \langle O (t) \rangle_{reg} &\equiv& \langle0| O (t) |0\rangle - \langle0_t| O (t) |0_t\rangle\nn\\
&=& \langle0| O (t) - \tilde{O} (t) |0\rangle.\ea
In the second line, we introduced an operator $\tilde{O} (t)$ in which all the fields are defined after Bogoliubov transformations. We follow \cite{Enqvist:2012im} where it was proposed, for an operator formed by products of several bilinear spinor fields, that the regularized operator can be written by simply dressing the wave functions by the occupation number,
\be 
\tilde{u}^{(|\lambda|)}_{\textbf{p},\pm}=\sqrt{2} |\beta_\textbf{p}^{(|\lambda|)}| u_{\textbf{p},\pm}^{(|\lambda|)}.
\ee
Through  the use of the regularized wave functions,
\be \tilde{W}^{(|\lambda|)}_{\textbf{k}, \textbf{p}} (t) = 2 |\beta^{(|\lambda|)}_\textbf{p} (t)| |\beta^{(|\lambda|)}_{\textbf{p}'} (t)| \, \{u_{\textbf{p}, +}^{(|\lambda|)}(t) u_{\textbf{p}', +}^{(|\lambda|)}(t) - u_{\textbf{p}, -}^{(|\lambda|)}(t) u_{\textbf{p}', -}^{(|\lambda|)}(t) \},\label{RegWF}\ee
we get an effective ultraviolet cutoff as particles are not excited when occupation numbers vanish.

For $\lambda = \pm\frac{1}{2}$, the mode decomposition (\ref{solRS2}) is more involved,
\be 
\tilde{\psi}_{\textbf{p}, \pm\frac{1}{2}}^{\mu} (t) = \sqrt{\frac{2}{3}}\epsilon_{\textbf{p}, 0}^\mu \,\mathbf{u}^{(\frac{1}{2})}_{\textbf{p} ,\pm\frac{1}{2}} (t) + \sqrt{\frac{1}{3}}\epsilon^\mu_{\textbf{p}, \pm1}\, \mathbf{u}^{(\frac{1}{2})}_{\textbf{p}, \mp\frac{1}{2}} (t).
\ee
Since  both $\epsilon_{\textbf{p}, 0}^\mu$ and $\epsilon^\mu_{\textbf{p}, \pm1}$ appear, we have $2^4 = 16$ helicity combinations in the four-point correlation functions. In the relativistic limit $p \gg m_{3/2}$, one could expand $\epsilon_{\textbf{p}, 0}^\mu$,
\be \epsilon_{\textbf{p}, 0}^\mu = \frac{1}{m_{3/2}} (p, \sqrt{p^2 + m_{3/2}^2} \hat{\textbf{p}}) = \frac{p^\mu}{m_{3/2}} + \frac{m_{3/2}}{2p} (-1, \hat{\textbf{p}}) + O (\frac{m_{3/2}^2}{p^2}).\ee
Thus, in the relativistic regime, we expect the leading order result to be obtained by replacing $\epsilon_{\textbf{p}, 0}^\mu \rightarrow \frac{p^\mu}{m_{3/2}}$. However, correlators with the four $\epsilon_{\textbf{p}, 0}^\mu$ inside Eq. (\ref{UTC4spinors}) replaced by $\frac{p^\mu}{m_{3/2}}$ vanish. The dominant contribution comes then from terms in  in Eq. (\ref{UTC4spinors}) where two of the four $\epsilon_{\textbf{p}, r}^\mu$ are $\epsilon_{\textbf{p}, 0}^\mu$. Notice that the two $\epsilon_{\textbf{p}, 0}^\mu$ can't be inside the same ${\hat \Pi^{lm}}(\textbf{p}, t)$ since the leading order of the stress tensor in Eq. (\ref{PiijTT22}) would vanish. Thus
there are $\begin{pmatrix}4 \\ 2 \end{pmatrix} -2 = 4$ such helicity combinations in all, which lead to
\ba 
\Pi^2_{\frac{1}{2}} (k, t, t') \simeq \frac{1}{2 \pi^2} \int_{p, p' \gg m_{3/2}} dp\, d\theta \, \, \, \,  K^{(\frac{1}{2})}(p, k, \theta, m_{3/2}) \, \, \, \tilde{W}^{(\frac{1}{2})}_{\textbf{k}, \textbf{p}} (t) \tilde{W}^{(\frac{1}{2})*}_{\textbf{k}, \textbf{p}} (t') \, .
\label{UTC1/2}
\ea
\newpage
Here, we defined
\ba 
K^{(\frac{1}{2})}(p, k, \theta, m_{3/2}) &= &
\frac{1}{36 m_{3/2}^2} p^4p'^2 \sin\theta \{ ({\cos}\theta - {\cos}\theta')^2 + 4 \sin^4 (\frac{\theta-\theta'}{2}) (1 + \sin\theta \sin\theta')\}  \nonumber\\
 &+&   \cdots,
 \label{defK}
 \ea
where we omit the terms subleading and proportional to $m_{3/2}$ in $\cdots$. 

Note that the factors $p^2/m_{3/2}^2$  are those expected for longitudinal modes following the equivalence theorem. The integral in Eq. (\ref{UTC1/2}) has only contributions from the relativistic regime where the equivalence theorem for spin-3/2 massive states shows that the couplings of their helicity-1/2 components are enhanced with respect to the helicity-3/2 ones \cite{Fayet:1977vd, Casalbuoni:1988qd, Maroto:1999vd} by  factors of  $p/m_{3/2}$. We expect the helicity-1/2 components to produce stronger  gravitational wave signals. One can compare $K^{(\frac{1}{2})}$ (\ref{defK}) to $K^{(\frac{3}{2})}$ (\ref{defK32}) and the dependence for spin-1/2 fermions. The latter was found in \cite{Enqvist:2012im} to scale like $p^4 \sin^3 \theta$.

It may be easier to understand the equivalence theorem when the spin-3/2, here the gravitino, acquires a mass through a super-Higgs mechanism. Imposing cancellation of the vacuum energy allows one to identify the scale of supersymmetry breaking as  $\sqrt{F} = \sqrt{\sqrt{3}m_{3/2}M_{Pl}}$. The power law behavior is then valid for momenta in the range $m_{3/2} \ll p \ll  \sqrt{\sqrt{3}m_{3/2}M_{Pl}}$.  Discussion of the necessity of this UV cutoff  using a bottom-up approach for  massive Rarita-Schwinger fields minimally coupled to gravity can also be found for example in\cite{Rahman:2011ik}. Therefore, in our computation the maximum energy scale for $p$ and $p'$, which corresponds to the vanishing occupation number through the regularization process Eq. (\ref{RegWF}), is required to be below this cutoff. Indeed, in the example of the next section, we would see that the nonadiabatic production of fermions forms a Fermi sphere whose radius $k_F$ is related to the mass of the scalar field source, and thus to the symmetry breaking scale.

Finally, we plug (\ref{UTC1/2}) into the subhorizon spectrum  (\ref{ gravitational wavespectrum}).  Taking into account the background evolution, we get
\be 
\frac{d\rho_{ GW}}{d \textrm{log} k} (k, t) \simeq \frac{G k^3}{\pi^3 a^4 (t)} \int dp\, d\theta \,  K^{(\frac{1}{2})}(p, k, \theta, m_{3/2})\ \{| I_{c} (k, p, \theta, t) |^2 + | I_{s} (k, p, \theta, t) |^2\},
\label{ GWspin-3/2 state}
\ee
where
\be 
I_{c} (k, p, \theta, t) = \int_{t_i}^t \frac{dt'}{a(t')} {\cos} (kt') \,\tilde{W}^{(\frac{1}{2})}_{\textbf{k}, \textbf{p}} (t'), \qquad I_{s} (k, p, \theta, t) = \int_{t_i}^t \frac{dt'}{a(t')} \sin (kt') \,\tilde{W}^{(\frac{1}{2})}_{\textbf{k}, \textbf{p}} (t')
\label{spin-3/2 statespectrum}
\ee
parametrize the spectrum of helicity-1/2 component. Then, (\ref{ GWspin-3/2 state}) is the master equation for gravitational waves produced from nonadiabatic spin-3/2 gases.

\section{Spin-3/2 state produced during preheating and  gravitational wave spectrum}

The knowledge of the wave functions $u^{(|\lambda|)}_{\textbf{p}, \pm}$ is necessary in order to extract quantitative results  for the expected gravitational waves spectrum. We consider here the production of gravitinos during preheating in order to illustrate our results through a simple but explicit toy model example.

Processes that produce gravitinos in the early Universe can be separated in two classes: thermal and nonthermal. Unless they are more massive than the reheat temperature, gravitinos can always be thermally produced in scattering of the particles in the thermal bath  \cite{Moroi:1993mb}. A nonthermal case is provided for example when gravitinos are produced during preheating  by the conversion of the energy stored in the coherently oscillating scalar field. As this process is nonadiabatic, it provides a possible framework for production of gravitational waves. For our purpose, we consider the case where supersymmetry breaking by the $F$ terms always dominates the one from the inflaton (i.e., the curvature). This ensures that 
$\mathcal{H} \ll m_{3/2}$, which allows one to neglect curvature when discussing the production of gravitational waves as we have assumed in the previous section.

We consider the case of generation of gravitational waves by longitudinal modes of gravitinos. In a FRW background, the corresponding equations of motion can be written in the form \cite{Kallosh:1999jj, Giudice:1999yt}
\begin{equation}
    [i\gamma^0 \partial_0 - a \, m_{3/2} + (A+iB\gamma^0) \textbf{p} \cdot \mathbf{\gamma}]\begin{pmatrix}u_+ \\ u_- \end{pmatrix} = 0,
\label{eqMotionG}
\end{equation}
where we have omitted the label $\frac{1}{2}$ and $a$ is the conformal factor in (\ref{FRW}). The functions $ A $ and $B$ satisfy $ A^2+ B^2 = 1$. 

As the initial condition, $u^{(|\lambda|)}_{\textbf{p}, \pm}$ satisfies the vanishing occupation number condition in Eq. ({\ref{occupationnumber}}). Since the wave function is isotropic, without losing generality, we take the momentum $\textbf{p}$ to lie along the z direction. Following \cite{Giudice:1999yt}, we define
 \begin{equation}
 A+iB = \textrm{exp}(2i\int\theta(t)dt), \qquad {\rm and} \qquad f(t)_\pm = \textrm{exp}(\mp i\int\theta(t)dt) u_\pm \, ,
   \label{def equations of motionf}
   \end{equation}
and the equation of motion (\ref{eqMotionG}) becomes
 \begin{equation}
    \ddot{f}_\pm + [p^2+ (\theta + m_{3/2}a)^2 \pm i(\dot{\theta} +\dot{m_{3/2}a})]f_\pm = 0.
     \label{equations of motionf}
 \end{equation}

A simple example for the production of longitudinal modes from the oscillation of a scalar field is the Polonyi model  studied in \cite{Ema:2016oxl}.  The corresponding K\"{a}hler potential and superpotential were chosen to be
\ba 
\mathcal{K} &=& |z|^2 - \frac{|z|^4}{\Lambda^2},\\
\mathcal{W} &=& \mu^2 z + \mathcal{W}_0,
\ea
where $z$ is the Polonyi field. An estimate of the mass order near the minimum is
\be 
m_{3/2} \simeq \frac{\mu^2}{\sqrt{3}M_{Pl}^2} \simeq \frac{\mathcal{W}_0}{M_{Pl}^2}, \qquad m_z \simeq 2\sqrt{3} \frac{m_{3/2} M_{Pl}}{\Lambda}. 
\ee
Here $\simeq$ means, in particular, that we neglect the higher order terms suppressed by $M_{Pl}$ and tune the cosmological constant to almost $0$. Requiring $\Lambda \ll M_{Pl}$ leads to $m_z \gg m_{3/2}$. It is also assumed that the $F$ term of $z$ does not contribute to the Hubble expansion but is large enough to lead to a gravitino mass that satisfies
\be 
\mathcal{H} \ll m_{3/2} \ll  m_z,
\ee 
in agreement with previous section assumptions; in particular, the background curvature can be neglected and we can use the flat space quantization. 
It was shown in \cite{Ema:2016oxl}, even in this limit, that the Polonyi model contains a nontrivial source term  $\theta (t)$ in Eq. (\ref{equations of motionf}) to produce helicity-1/2 gravitino,
\begin{equation}\theta(t) = -\frac{a{m^2_z} {\delta z}}{2\sqrt{3}m_{3/2}M_{Pl}} = -\frac{a{m^2_z} {\delta z}}{2 F},\end{equation}
where $\delta z = z-z_0$ is the displacement of $z$ from its value $z_0$ at the minimum of the the scalar potential and $F = \sqrt{3}m_{3/2}M_{Pl}$ is the supersymmetry breaking scale.

\begin{figure}[H]
 		\centering
 		\includegraphics[width=0.80\columnwidth]{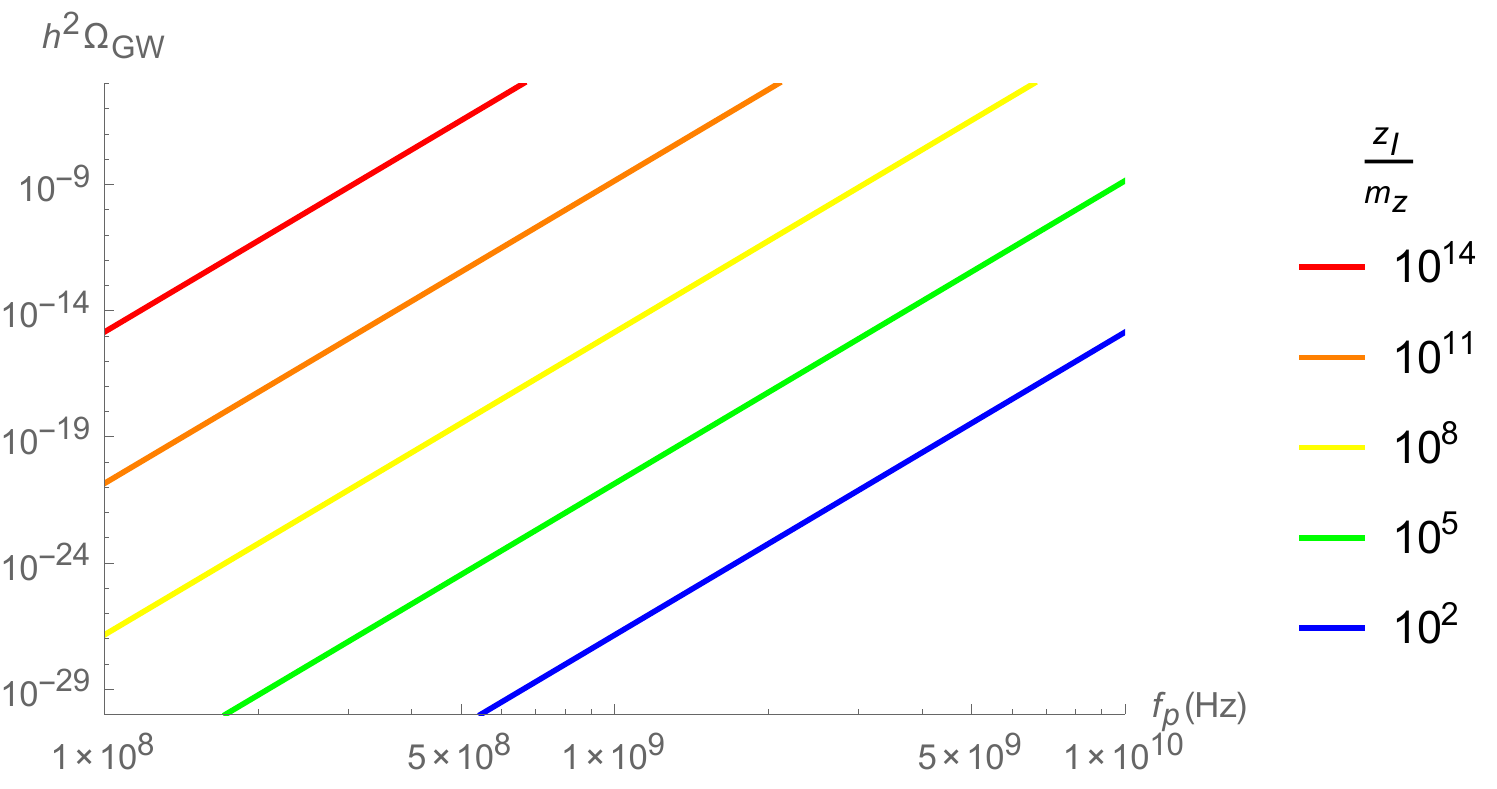}
 		\caption{
 			\footnotesize
 			{The peak amplitude of  gravitational wave according to Eq. (\ref{PA}).}
 		}
 		\label{PAplot}
 	\end{figure}

In order to estimate the production of longitudinal modes of the gravitino,  we need  their couplings to $\delta z$.  This is contained in the corresponding equation of motion (\ref{equations of motionf}),
\begin{equation}
   \ddot{f}_\pm + [k^2+ (a \,m_{3/2} -\frac{a{m^2_z} {\delta z}}{2 F})^2 \mp i \frac{a{m^2_z} {\dot{\delta z}}}{2 F}]f_\pm = 0.
   \label{equations of motion1/2production}
\end{equation}
We can see that (\ref{equations of motion1/2production}) for helicity-1/2 gravitino resembles the form of that of spin-1/2 fermions produced nonadiabatically from Yukawa coupling with a coherently oscillating scalar in the quadratic potential.  Thus one expects the spectrum of helicity-1/2 gravitino and Eq. (\ref{spin-3/2 statespectrum}) in this model to be similar to the spin-1/2 fermion cases considered in \cite{Enqvist:2012im}. The effective Yukawa coupling  $\tilde{y}$, required to be smaller than 1 by unitarity, reads
\begin{equation}
       \tilde{y} = \frac{m_z^2}{2F}.
\label{ytilde}\end{equation} 
According to \cite{Greene:1998nh}, the fermion production in this case is expected to fill up a Fermi sphere with comoving radius,
\be 
k_F \sim (a/a_I)^{1/4}q^{1/4}m_z, \qquad q \equiv \frac{\tilde{y}^2 z_I^2}{m_z^2}, 
\ee
where $q$ is the resonance parameter and $z_I$ is the initial vacuum expectation value (i.e., where inflation ends) of the Polonyi field. Outside the Fermi sphere, the occupation number decreases exponentially. Thus one can expect that the peak of the  gravitational wave spectrum corresponds to the radius of Fermi-sphere $k_p \sim k_F$, which in the present leads to the characteristic frequency,
\be 
f_p\simeq 6\cdot10^{10} \tilde{y}^{\frac{1}{2}} \textrm{Hz}.
\ee
One can see from above and Eq. (\ref{ytilde}) that the validity of the effective field theory requires the peak frequency to be below $10^{10}$ Hz. The amplitude at the peak of the gravitational wave spectrum can also be estimated by taking the result of a spin 1/2 field and multiplying it by the enhancement factor $(\frac{k_F}{m_{3/2}})^2$,
\ba 
h^2\Omega_{ GW} (f_p) &\simeq& 2.5 \cdot 10^{-12} (\frac{m_z^2}{z_I M_{Pl}})^2(\frac{a_*}{a_I})^{\frac{1}{2}} q^{\frac{3}{2}} (\frac{k_F}{m_{3/2}})^2 \nn \\
&=& 3 \cdot 10^{-11} \tilde{y}^6 (\frac{z_I}{m_z})^2 (\frac{a_*}{a_I}) \nn \\
&\simeq& 3 \cdot 10^{-10} (\frac{f_p}{6\cdot 10^{10} \textrm{Hz}})^{12} (\frac{z_I}{m_z})^2,
\label{PA}
\ea
where $a_I$ and $a_*$ are the scale factor at initial time and the end of  gravitational wave production. We assumed an order 10 increase for scale factor in the last line. The relation between the amplitude and the peak frequency is shown in Fig. (\ref{PAplot}). If we take  $z_I = 10^{-3}\,M_{Pl}, m_z = 10^{10}$ GeV, we get that the amplitude at peak frequency $3 \cdot 10^9$ Hz is $7.3 \cdot 10^{-16}$. For a peak appearing at lower frequency, the amplitude is too small to be observed due to the power $12$ in Eq. (\ref{PA}). For very large value of $\frac{z_I}{m_z}$, one can consider that the low frequency tail gets enhanced enough to become observable at lower frequency detector.
Near the peak, the  dominant part of $K^{(\frac{1}{2})}$  (\ref{defK}) scales as $p^4 \frac{k^2}{m_{3/2}^2}$, which makes the spectrum  (\ref{ GWspin-3/2 state}) go as $k^5$. In the lower frequency, the spectrum becomes $k^3$, thus scaling again like the spin $1/2$. These two different scaling behaviors are a feature of gravitational wave spectrum produced from spin-3/2 particles, which is quite model independent due to the presence of a filled-up Fermi sphere.

\section{Conclusion}

The  gravitational wave signals from spin-3/2 fermions are especially interesting since the latter are the only missing piece in the nature with spin between 0 and 2. Moreover, their presence can be a smoking gun for supersymmetry playing a role in the early Universe. The nonadiabatic production of helicity-1/2 gravitino takes the similar form as spin-1/2 fermions nonadiabatically produced from coherently oscillating scalars with quadratic potential. Thus it fills up a Fermi sphere in the occupation number. The corresponding comoving radius governs the position of the peak frequency of the  gravitational waves. Their spectrum has two main differences compared to the one from spin-1/2 fermions. First is that the order of the amplitude gets enhanced by a factor of $\frac{k_F^2}{m_{3/2}^2}$. Second, the stress tensor of spin-3/2 state contains a term proportional to the  gravitational wave mode $\textbf{k}$, while for scalars and spin-1/2 fermions, the $\textbf{k}$ dependence is projected out by the projector $\Lambda_{ij,lm} (\textbf{k})$ on traceless-transverse modes. Thus one could expect a $k^5$ dependence for the  gravitational wave spectrum near the peak. The observed window for the gravitational waves, like most preheating scenarios, lies at very high frequency, around $10^{9}$ Hz in our simple example, which calls for the design of new experiments like \cite{Sabin:2014bua}.  We leave for future work investigating further examples.

\section*{Acknowledgments}
 We are grateful to  I.~Antoniadis, W.~Buchm\"uller, L.~Darm\'e, V.~Domcke, Y.~Ema, W.~Huang, S.~Patil, F.~Sala, and T.~Terada for useful discussions. We acknowledge the support of  the Agence Nationale de Recherche under Grant No. ANR-15-CE31-0002 ``HiggsAutomator''. This work is also supported by the Labex ``Institut Lagrange de Paris'' (Grants No. ANR-11-IDEX-0004-02, and No.  ANR-10-LABX-63).

\end{document}